\documentclass[lettersize,journal]{IEEEtran}
\usepackage{amsmath,amsfonts}
\usepackage{algorithmic}
\usepackage{algorithm}
\usepackage{array}
\usepackage[caption=false,font=normalsize,labelfont=sf,textfont=sf]{subfig}
\usepackage{textcomp}
\usepackage{stfloats}
\usepackage{url}
\usepackage{verbatim}
\usepackage{graphicx}
\usepackage{cite}
\usepackage{booktabs}  
\usepackage{multirow}  
\usepackage{pgfplots}
\usepackage[hidelinks]{hyperref}
\usepackage{orcidlink}
\pgfplotsset{compat=1.18}
\begin{document}

\title{Lossless Inference Acceleration for ASR via EAGLE-Style Speculative Decoding}

\author{Sen Zhang, Xianghu Yue, Yiming Chen, Wenhuan Lu, Jianguo Wei,~\IEEEmembership{Senior Member,~IEEE}, \\ Haizhou Li,~\IEEEmembership{Fellow,~IEEE}

\thanks{Sen Zhang, Xianghu Yue, Wenhuan Lu, and Jianguo Wei are with the College of Intelligence and Computing, Tianjin University, Tianjin 300350, China (email: zhangsen@tju.edu.cn; yuexianghu@tju.edu.cn; wenhuan@tju.edu.cn; jianguo@tju.edu.cn)}
\thanks{Yiming Chen is with ASUS Intelligent Cloud Services, Singapore 486030 (email: mattymchen@gmail.com).}
\thanks{Haizhou Li is with the School of Artificial Intelligence, The Chinese University of Hong Kong, Shenzhen, Guangdong 518172, China, also with the Department of Electrical and Computer Engineering, National University of Singapore, Singapore 117583 (email: haizhouli@cuhk.edu.cn).
}}

\maketitle

\begin{abstract}
Autoregressive Transformer and Conformer models have achieved state-of-the-art performance in automatic speech recognition (ASR).
However, their sequential token generation mechanism imposes a significant bottleneck during decoding, resulting in substantial inference latency.
While speculative decoding has emerged as a promising lossless acceleration technique, existing ASR-oriented approaches either suffer from low draft acceptance rates due to suboptimal draft quality or rely on specially designed external draft models, which limit their generality and practicality.
To overcome these limitations, we propose EAGLE-ASR, a novel framework that accelerates ASR inference via EAGLE-style speculative decoding while strictly preserving lossless output quality. 
Specifically, we introduce Speech-Draft, a lightweight autoregressive module that serves as a universal draft model by jointly leveraging encoder acoustic features and hierarchical decoder representations from the target ASR model. 
Furthermore, we adopt a multi-step generation strategy with self-distilled data to train the Speech-Draft, enabling it to better approximate the target model's output distribution and maintain causal consistency during multi-token generation. 
Extensive experiments on representative Whisper and Conformer AED models demonstrate that EAGLE-ASR achieves significant lossless speedups of up to 1.9$\times$ and 2.4$\times$, respectively, consistently outperforming existing speculative decoding baselines for ASR. 
Our proposed framework further generalizes effectively to multilingual ASR and speech translation tasks, establishing EAGLE-ASR as a practical and universally applicable solution for accelerating modern ASR systems without compromising recognition accuracy.
Our code is publicly available at \href{https://github.com/ssss-sen/eagle-asr}{https://github.com/ssss-sen/eagle-asr}.
\end{abstract}


\section{Introduction}
\label{sec:intro}
\IEEEPARstart{N}{eural} network models based on Transformer or Conformer architectures have recently achieved state-of-the-art performance in automatic speech recognition (ASR)~\cite{radford2023robust, gulati2020conformer, baevski2020wav2vec, vaswani2017attention}. 
However, these accuracy improvements in transcription are often accompanied by increasingly large model sizes and more complex decoding procedures.
In particular, autoregressive ASR models generate output tokens sequentially, which bottlenecks the decoding stage and introduces substantial inference latency. As ASR systems are increasingly deployed on resource-constrained edge devices and in high-throughput server environments, reducing per-utterance decoding latency without sacrificing recognition accuracy has become a critical challenge for the speech processing community~\cite{prabhavalkar2023end}, \cite{gondi2021performance}, \cite{nayeem2025automatic}.

Streaming ASR techniques have demonstrated effectiveness in reducing end-to-end latency by emitting partial results before processing the entire audio utterance\cite{graves2012sequence, he2019streaming, shi2021emformer, zhang2020transformer}.
In this work, however, we focus on accelerating the decoding stage of autoregressive ASR models under the standard full-utterance inference setting, which remains crucial for offline transcription.
To improve decoding efficiency in this setting, various acceleration strategies have been explored, including model compression~\cite{gandhi2023distil, shao2024dq, gu2025efficient, kamahori2025liteasr} and non-autoregressive (NAR) decoding \cite{higuchi2020mask, gao2022paraformer, an2024paraformer, chi2021align, chen2020non}. Model compression techniques aim to reduce computational cost and memory footprint by constructing smaller or more efficient variants of large-scale models. On the other hand, NAR decoding attempts to mitigate the latency of sequential generation by predicting multiple tokens in parallel.
Although these methods can reduce inference cost, they often require architectural modification or may sacrifice recognition accuracy for decoding speed. 
Therefore, lossless acceleration techniques for autoregressive ASR models remain highly desirable.

Speculative decoding has recently emerged as an effective inference acceleration technique for autoregressive models in natural language processing (NLP)\cite{leviathan2023fast}, \cite{chen2023accelerating}, \cite{xia2024unlocking}. 
It typically uses a lightweight draft model to generate multiple candidate tokens, which are then verified in parallel by the target model.
By reducing the number of expensive target-model decoding steps, speculative decoding can substantially improve inference speed while preserving the output distribution of the original model.

Recent studies have explored speculative decoding for ASR acceleration, but face critical limitations.
Some approaches~\cite{gandhi2023distil, wei2025specasr} rely on specially designed external draft models that must be well aligned with the target ASR model, increasing implementation complexity and limiting generalizability.
Others~\cite{segal2025whisper}, \cite{ho2025model} generate low-quality drafts, resulting in poor acceptance rates that constrain achievable speedups. 
These limitations motivate the need for an ASR-oriented speculative decoding framework that generates high-quality drafts while remaining lightweight and adaptable.
EAGLE-style speculative decoding constructs lightweight draft modules by leveraging the target model's internal representations, thereby eliminating the need for external models~\cite{li2024eagle, li2024eagle2, li2025eagle, cai2025fastmtp}. While this design is highly attractive for ASR, its direct application poses non-trivial challenges. Unlike pure text generation tasks, ASR involves cross-modal conditioning on continuous acoustic features, where the decoder's behavior is inherently coupled with encoder representations. A naive adaptation that ignores acoustic context or decoder–encoder interactions can result in poor draft quality and reduced acceptance rates, ultimately undermining the efficiency gains of speculative decoding~\cite{saon2026self}, \cite{huang2025specvlm}, \cite{kang2026vispec}.

To address these challenges, we propose EAGLE-ASR, a speculative decoding framework tailored to autoregressive ASR models. The key component of EAGLE-ASR is Speech-Draft, a lightweight autoregressive draft module that leverages both encoder-side acoustic features and hierarchical decoder representations from the target ASR model. 
By incorporating acoustic and decoder-context information, Speech-Draft better approximates the target model’s token prediction behavior while maintaining low computational cost, thereby improving draft acceptance and inference efficiency.
Furthermore, we design a self-distillation and multi-step generation strategy to train Speech-Draft. 
This training strategy enables the draft module to model dependencies among consecutive future tokens and maintain causal consistency during multi-token drafting. 
As a result, EAGLE-ASR can generate multiple high-quality draft tokens at each decoding step and verify them efficiently with the target ASR model.
We evaluate EAGLE-ASR on representative Whisper and Conformer attention-based encoder-decoder (AED) models. 
Experimental results show that EAGLE-ASR achieves average accepted draft lengths of 3.67 on Whisper and 4.74 on the Conformer AED model, demonstrating effective draft generation. 
It further achieves lossless inference speedups of up to 1.9$\times$ and 2.4$\times$, respectively, outperforming existing ASR speculative decoding baselines. 
Additional experiments also demonstrate the effectiveness of the proposed framework on multilingual ASR and speech translation tasks.


The main contribution of this paper can be summarized as follows.

\begin{itemize}
\item We propose EAGLE-ASR, an EAGLE-style speculative decoding framework for accelerating autoregressive ASR inference while preserving recognition accuracy.
\item We introduce Speech-Draft, a lightweight autoregressive draft module that leverages encoder acoustic features and hierarchical decoder representations to generate high-quality draft tokens for attention-based encoder-decoder ASR models.
\item We develop a self-distillation and multi-step generation training strategy that improves draft quality and maintains causal consistency during multi-token speculative decoding.
\item Extensive experiments on Whisper and Conformer AED models demonstrate that EAGLE-ASR achieves significant lossless speedups of up to 1.9$\times$ and 2.4$\times$, respectively, consistently outperforming existing ASR speculative decoding baselines.
\end{itemize}

The remainder of this paper is organized as follows: Section \ref{sec:works} summarizes related work. Section \ref{sec:method} presents the architecture of Speech-Draft, along with the training and inference framework of EAGLE-ASR. Section \ref{sec:experiment} describes the experimental setup, with results provided in Section \ref{sec:result}. The paper concludes in Section \ref{sec:conclude}.

\section{Related Work}
\label{sec:works}
\subsection{Traditional Inference Acceleration Techniques in ASR}
Traditional approaches for accelerating ASR inference mainly focus on model compression and non-autoregressive ASR (NAR-ASR). 
Model compression techniques, such as knowledge distillation\cite{hinton2015distilling}, quantization\cite{gholami2022survey}, and pruning\cite{han2015learning}, reduce computational overhead by decreasing model size and complexity. Representative works, including Distil-Whisper\cite{gandhi2023distil} and DQ-Whisper\cite{shao2024dq} achieve faster inference by using compressed model variants of large ASR models. 
However, these approaches often introduce some degree of information loss, which may lead to degradation in Word Error Rate (WER). Concurrently, NAR-ASR models\cite{gu2018non} attempt to eliminate the latency inherent in autoregressive decoding by generating multiple tokens in parallel. Notable approaches include Mask-CTC\cite{higuchi2020mask}, Paraformer\cite{gao2022paraformer}, \cite{an2024paraformer}, Align-refine\cite{chi2021align} and A-CMLM\cite{chen2020non} models. Despite their improved decoding efficiency, NAR-ASR approaches generally exhibit a performance gap compared to autoregressive models in terms of recognition accuracy. Overall, while these strategies demonstrate effectiveness in reducing inference latency, they typically sacrifice recognition performance for speed. This limitation highlights the need for lossless acceleration methods that are specifically tailored to autoregressive ASR models.

\subsection{Speculative Decoding}
Speculative decoding is a lossless acceleration technique for large language models that employs a ``draft-then-verify'' paradigm. Early speculative decoding methods\cite{stern2018blockwise}, \cite{sun2021instantaneous} accelerated generation in greedy settings, while recent advances\cite{leviathan2023fast, chen2023accelerating} introduced speculative sampling to extend this framework to non-greedy generation. In this paradigm, the drafting phase incurs minimal computational overhead, while the verification phase is executed in parallel, corresponding to draft generation and validation, respectively. In the drafting phase, a lightweight draft model is employed to autoregressively generate tokens, forming a draft sequence while recording the probability $\hat{p}$ for each token. In the verification phase, the target model is invoked to evaluate the draft and record the probability $p$ for each drafted token. Speculative decoding then sequentially determines the acceptance of each drafted token with an acceptance probability of $\min(1, p/\hat{p})$. If the current token is accepted, the process proceeds to the next token; otherwise, a replacement token is sampled from the distribution $\text{norm}(\max(0, p - \hat{p}))$, and subsequent tokens in the draft are discarded\cite{miao2024specinfer}, \cite{spector2023accelerating}. Crucially, Leviathan et al.\cite{leviathan2023fast} proved that speculative decoding is consistent with the distribution of vanilla autoregressive decoding. Building upon this foundation, numerous variants have emerged in the NLP domain, including CS Drafting\cite{chen2024cascade}, BiLD\cite{kim2023speculative}, LLMA\cite{yang2023inference}, and REST\cite{he2024rest}, each exploring different strategies for draft generation and verification.

Unlike conventional speculative decoding approaches that rely on separate, independent draft models, Medusa\cite{cai2024medusa} and EAGLE\cite{li2024eagle}, \cite{li2024eagle2}, \cite{li2025eagle} adopt an internally integrated speculative decoding strategy, which aims to directly leverage the feature extraction capability of the target model itself for draft generation.

Specifically, Medusa attaches multiple lightweight MLP heads on top of the frozen target model, utilizing the hidden states at the current time step to predict candidate tokens for the subsequent $k$ positions in parallel. However, because these heads predict each position independently, the approach fails to capture the sequential dependencies among the drafted tokens. Consequently, each later token is generated without conditioning on its immediate predecessor, resulting in less coherent draft sequences and a diminished acceptance rate during the verification phase.

In contrast to Medusa's multi-head design, EAGLE attaches only a single lightweight autoregressive module to the frozen target model, leveraging semantic information from different layers of the target model for prediction. By maintaining an explicit autoregressive structure, this design preserves temporal dependencies during draft generation while more effectively leveraging the target model’s learned representations. Empirically, EAGLE has been reported to achieve state-of-the-art decoding speed among existing speculative decoding approaches.

\subsection{Speculative Decoding in ASR}
Current research on accelerating ASR models through speculative decoding still largely focuses on selecting or training a separate draft model. For instance, the Hugging Face team employs Distil-Whisper and Whisper-tiny as draft models to accelerate the inference of the Whisper-large target model\cite{gandhi2023distil}. However, such approaches rely heavily on strict distribution alignment between the draft and target models, typically requiring them to belong to the same model family or the draft model to be a distilled variant of the target. This constraint significantly limits architectural flexibility and practical deployment. When this assumption is violated, such as when pairing a Distil-Whisper draft model with a Conformer-based target model, the token acceptance rate can degrade substantially, thereby diminishing or even negating the anticipated speedup.

Whisper-Medusa partially alleviates the cost of obtaining a well-aligned draft model by appending multiple MLP heads on top of a Whisper model\cite{segal2025whisper}. However, constrained by the inherent limitations of the Medusa framework, this approach similarly neglects temporal dependencies and underutilizes the rich intermediate representations extracted by the target model, particularly the acoustic features derived from the encoder. Consequently, its overall acceleration ratio remains limited. Moreover, since Whisper-Medusa fine-tunes the final decoder layer during training rather than freezing all parameters of the target model, its WER exhibits noticeable degradation, thus failing to achieve lossless acceleration.

To address these limitations, we propose EAGLE-ASR, an EAGLE-style speculative decoding framework tailored for accelerating the inference of the ASR model. Unlike existing approaches, EAGLE-ASR eliminates the need for an independently trained draft model while delivering superior decoding speedups and strictly preserving lossless output quality.

\begin{figure*}[ht]
	\centering 
	\includegraphics[width=1\textwidth]{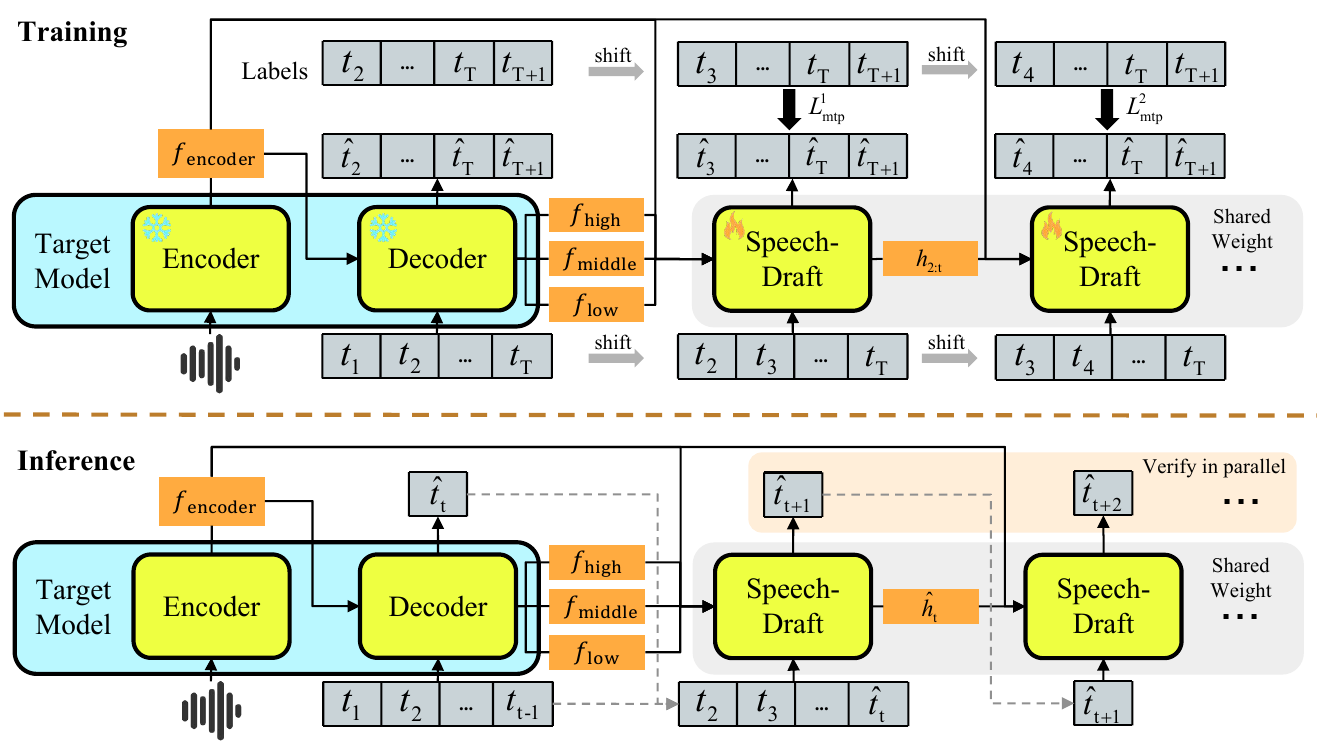} 
	\caption{Pipeline of EAGLE-ASR. {\bf Training phase (upper): }The Speech-Draft model is trained using a multi-token generation strategy together with a weighted cross-entropy loss with exponential decay. During training, only the FC layers and the Transformer decoder block of the Speech-Draft are fine-tuned, while remaining parameters of the Speech-Draft, as well as all parameters of the target model, are kept frozen. {\bf Inference phase (lower): }The target model predicts the next token, which feeds into the Speech-Draft serving as the draft model, thereby recursively generating draft tokens for subsequent parallel verification.} 
	\label{eagle-asr} 
\end{figure*}

\section{Methodology}
\label{sec:method}

\subsection{Overview}
\label{infer}
In this work, we propose EAGLE-ASR, an EAGLE-style speculative decoding framework designed to accelerate autoregressive ASR inference while maintaining lossless recognition accuracy.
As illustrated in \autoref{eagle-asr}, EAGLE-ASR is built upon a frozen target ASR model consisting of an encoder and an autoregressive decoder, augmented with a lightweight autoregressive module termed Speech-Draft and a dedicated speculative decoding pipeline. 
Unlike conventional speculative decoding methods that rely on standalone external draft models, Speech-Draft directly leverages the internal representations of the target ASR model for draft generation, thereby improving target-draft consistency while maintaining low computational overhead.
Similar to standard speculative decoding, EAGLE-ASR alternates between a drafting stage and a verification stage during inference.
Specifically, given the input audio and the sequence prefix $t_{1:t-1}$, to generate multiple tokens in parallel, the inference process of EAGLE-ASR proceeds as follows, as depicted in the lower part of \autoref{eagle-asr}.

\textbf{Drafting stage:} During draft generation, the target ASR model first processes the input audio to predict the next token $\hat{t}_{t}$. Meanwhile, the encoder features $f_{\text{encoder}}$ are cached, and decoder representations at different levels,  denoted as $f_{\text{low}}$, $f_{\text{middle}}$, and $f_{\text{high}}$, are extracted from the target decoder. 
These multi-granularity features, together with the predicted token $\hat{t}_{t}$ and the sequence prefix $t_{1:t-1}$, are then fed into the Speech-Draft, which autoregressively generates $K$ draft tokens following the EAGLE framework~\cite{li2024eagle}.

\begin{itemize}
    \item In the initial draft step $k = 1$, the Speech-Draft takes as input $f_{\text{encoder}}$, $f_{\text{low}}$, $f_{\text{middle}}$, $f_{\text{high}}$, and the concatenated sequence $[t_{2:t-1}, \hat{t}_{t}]$, where the newly predicted token from the target model is appended to the input sequence shifted one step ahead. 
    Based on this information, Speech-Draft generates the first draft token $\hat{t}_{t+1}$.
    \item In the subsequent draft step $k > 1$, the Speech-Draft operates in an autoregressive manner. Specifically, it predicts the $k$-th draft token conditioned on the encoder representations $f_{\text{encoder}}$, the hidden state output from the previous drafting step $\hat{h}_{t+k-2}$, and the previously generated drafted token $\hat{t}_{t+k-1}$.
\end{itemize}

After $K$ recursive steps, a complete draft sequence $\hat{t}_{t+1:t+K}$ is obtained.

\textbf{Verification stage:} During verification, all drafted tokens are fed into the target decoder in parallel, which simultaneously computes the logits for positions $t+1$ through $t+K$. 
We adopt the standard speculative decoding acceptance criterion to determine the accepted draft tokens, where drafted tokens are accepted sequentially until the first position at which the draft token differs from what the target model would have sampled. 
Through this parallel verification without approximate strategies, Eagle-ASR ensures that the output distribution remains identical to that of the target model, thereby achieving acceleration in a lossless manner. 
In \autoref{hyperparameters}, we further investigate the impact of different verification strategies on acceleration performance. 
Experimental results show that linear verification yields a better speedup than tree-based verification in the ASR setting.

\subsection{Speech-Draft Module}
\label{speech-draft-arch}
\begin{figure}
\centering
\includegraphics[width=0.5\textwidth]{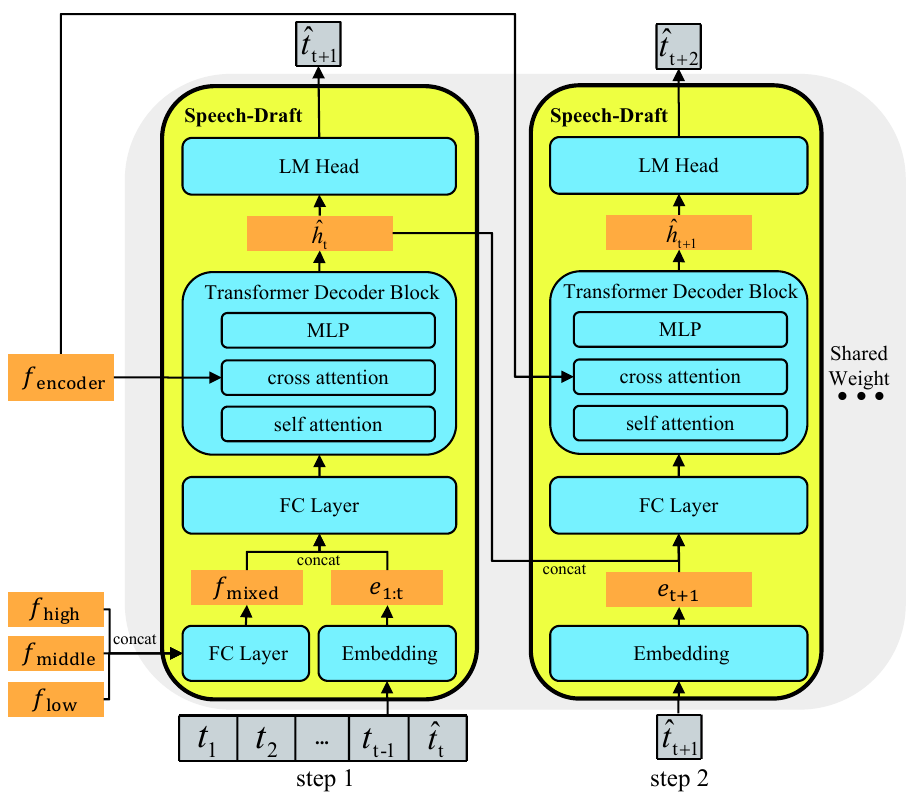}
\caption{The overall architecture of the Speech-Draft, which consists of two fully connected (FC) layers, an embedding layer, a Transformer decoder block, and an LM head.}
\label{speech-draft}
\end{figure}

Unlike standard language models, the autoregressive decoding in ASR exhibits a dual dependency, requiring both historical text context and acoustic representations produced by the encoder. 
Motivated by this observation, we propose Speech-Draft, a lightweight autoregressive module that effectively integrates acoustic and textual information to serve as the draft model in EAGLE-ASR. The overall architecture of the Speech-Draft is illustrated in \autoref{speech-draft}.
The Speech-Draft module consists of two fully connected (FC) layers, an embedding layer, a single Transformer decoder block, and an LM head. 
Notably, the embedding layer and the LM head directly reuse the corresponding parameters of the target model and require no additional training.

The first FC layer is activated only at the first decoding step ($\text{step}=1$), where it fuses the multi-level decoder representations ($f_{\text{low}}$, $f_{\text{middle}}$, and $f_{\text{high}}$) extracted from the target model into a unified representation $f_{\text{mixed}}$, thereby aggregating hierarchical information from different depths.
The second FC layer then combines $f_{\text{mixed}}$ (or the hidden states from the preceding step when $\text{step} > 1$)  with the embeddings of previously generated tokens, producing the input representation for the Transformer decoder block. 
To maintain strict alignment with the input speech signal, the encoder acoustic representation $f_{\text{encoder}}$ is incorporated into the drafting process through the cross-attention mechanism of the Transformer decoder block.
This design enables Speech-Draft to jointly leverage rich decoder-side semantic representations and fine-grained encoder-side acoustic information, thereby improving draft quality for subsequent verification.
It is worth noting that, owing to its lightweight and model-agnostic architecture, the Speech-Draft module can be seamlessly integrated into any autoregressive ASR system that employs a Transformer-based decoder, including state-of-the-art models such as Whisper and Conformer AED models.

\subsection{Self-Distilled Training Data}
\label{data-construct}
As illustrated in Figure 3, there is a distribution discrepancy between the output distribution of the target ASR model and the original distribution of the training data transcriptions.
This discrepancy arises from factors such as prediction errors, inaccurate annotations, and inconsistencies in transcription conventions.
Training Speech-Draft directly on the original transcriptions may therefore lead to suboptimal target-draft alignment, ultimately limiting speculative decoding performance.

\begin{figure}
\centering
\includegraphics[width=0.5\textwidth]{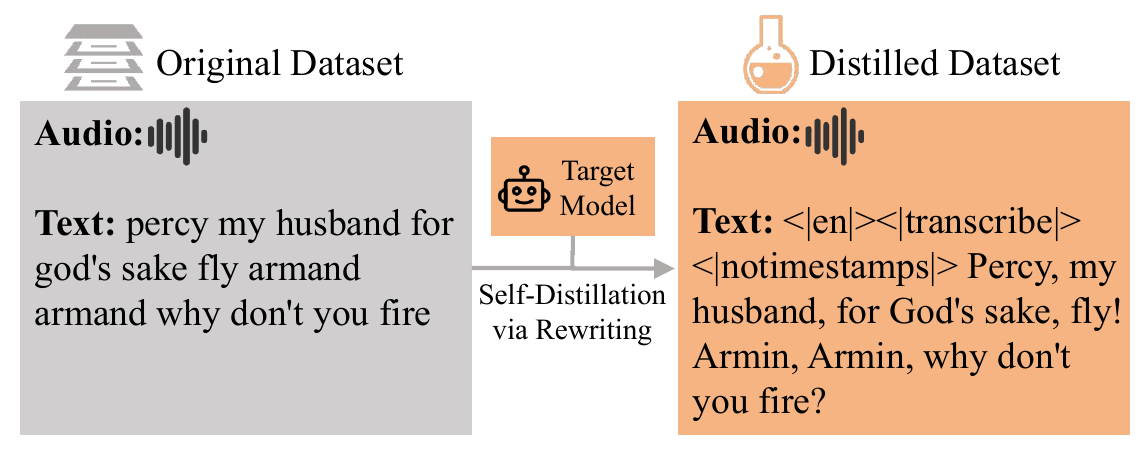}
\caption{An illustration of self-distilled training data generation, designed to mitigate the distributional discrepancy between the original transcripts and the target model outputs.}
\label{distiled-data}
\end{figure}

To address the above issue, we adopt a self-distillation approach for training data generation\cite{yang2024self}. 
Specifically, the target ASR model is first used to generate transcriptions for all training audio samples, and these generated transcriptions are subsequently used as the training targets for Speech-Draft.
For the Whisper model, we explicitly preserve its special tokens, such as \texttt{<|en|>}, \texttt{<|transcribe|>}, and \texttt{<|notimestamps|>}. 
The resulting self-distilled training data effectively captures the generation patterns and preferences of the target model.  
By training Speech-Draft on these self-distilled transcriptions rather than on the original annotations, the draft module learns to produce draft tokens that are consistent with the behavior of the target model, instead of attempting to imitate transcriptions from external sources that may exhibit a different distribution. 
As a result, Speech-Draft can better match the target model during speculative decoding, leading to a higher token acceptance rate and a more effective drafting process.

\subsection{Training Strategy}
\label{train-strategy}
During training, EAGLE-ASR updates only the trainable components of Speech-Draft, namely the FC layers and the Transformer decoder block.
The remaining parameters of Speech-Draft, including the reused embedding layer and LM head, as well as all parameters of the target ASR model, are kept frozen.
This strategy ensures computational efficiency during training while fully preserving the capabilities of the target model.

During actual speculative decoding, however, Speech-Draft faces a training–inference mismatch. Specifically, at the second and subsequent drafting steps, Speech-Draft must rely on its own predicted hidden states rather than the ground-truth hidden states produced by the target model.
This discrepancy can lead to error accumulation during recursive drafting, resulting in a degradation in acceptance rate. 
To alleviate this issue, EAGLE-ASR adopts a multi-step generation strategy for training Speech-Draft, as illustrated in the upper part of \autoref{eagle-asr}. 
Specifically, suppose that EAGLE-ASR employs Speech-Draft to perform $K$-step generation during training.
Given an input speech audio and its corresponding transcription token sequence $t_{1:T}$ (where $T$ denotes the sequence length), the training procedure of EAGLE-ASR is illustrated as follows.

\begin{itemize}
    \item At step $k = 1$, the Speech-Draft leverages $f_{\text{encoder}}$, $f_{\text{low}}$, $f_{\text{middle}}$, and $f_{\text{high}}$ from the target model, along with the shifted token sequence $t_{2:T}$ (shifted by one position), to predict $\hat{t}_{3:(T+1)}$.
    \item At step $k > 1$, the Speech-Draft utilizes $f_{\text{encoder}}$ and the output hidden states $\hat{h}_{k:T}$ from the preceding step, together with the shifted token sequence $t_{(k+1):T}$ (shifted by $k$ positions), to predict $\hat{t}_{(k+2):(T+1)}$.
\end{itemize}

As the generation depth increases, distant tokens condition on a growing number of intermediate decisions, resulting in progressively higher prediction difficulty. Consequently, the prediction uncertainty accumulates with each successive step. To mitigate this issue, we optimize the Speech-Draft using a weighted cross-entropy loss with exponential decay:

\begin{equation}
	{L}_{\text{mtp}} = \sum_{k=1}^{K} \alpha_k \cdot {L}_{\text{mtp}}^k = \sum_{k=1}^{K} \alpha_k \cdot \text{CE}\left( \hat{t}_{1+k:T+1}^k,\ t_{1+k:T+1} \right) \tag{1}
\end{equation}
where $L_{k}^{\text{mtp}}$ represents the loss at the $k$-th step, and $\text{CE}(\cdot, \cdot)$ denotes the cross-entropy loss between the predicted token and the ground-truth label. The position-dependent weight $\alpha_k$ follows an exponential decay, formulated as:

\begin{equation}
    \alpha_k = \frac{\beta^{k-1}}{\sum_{j=1}^{K} \beta^{j-1}}
    \tag{2}
\end{equation}
where $\beta$ denotes the decay factor. Through this loss formulation, EAGLE-ASR prioritizes the accuracy of near-term predictions while simultaneously enhancing its ability to generate multiple consecutive future tokens.

\section{Experimental Setup}
\label{sec:experiment}
\subsection{Datasets}
\label{datasets}
We use two datasets to train and evaluate the proposed EAGLE-ASR framework.
First, we adopt LibriSpeech, a widely used benchmark in ASR research\cite{panayotov2015librispeech}. LibriSpeech contains approximately 1,000 hours of read English speech derived from audiobooks in the public domain. 
In this work, we use the full 960-hour LibriSpeech training split to train the Speech-Draft module and report evaluation results on the standard test-clean and test-other subsets.
Moreover, to further evaluate the generalizability of EAGLE-ASR on both multilingual ASR and speech translation (ST) tasks, we conduct additional experiments on the VoxPopuli dataset\cite{wang2021voxpopuli}. VoxPopuli is a large-scale multilingual speech corpus derived from European Parliament event recordings between 2009 and 2020. 
It comprises approximately 400,000 hours of unlabeled speech data covering 23 languages, as well as around 1,800 hours of transcribed speech data in 16 languages. 
For multilingual ASR evaluation, we use the English, German, Spanish, French, and Italian subsets. 
For the ST task, we conduct experiments on four translation directions: German-to-English, Spanish-to-English, French-to-English, and Italian-to-English.

It is worth noting that, although both LibriSpeech and VoxPopuli subsets used in our experiments provide paired reference transcripts for all audio samples, we do not directly use these human-annotated transcriptions for training. 
Instead, we train the Speech-Draft exclusively on self-distilled data generated by the target model, as described in \autoref{data-construct}. 

\subsection{Model Configurations}
\label{models}
To validate the effectiveness and generality of EAGLE-ASR, we adopt two representative autoregressive ASR models, Whisper and Conformer AED, as our target models\cite{radford2023robust}, \cite{gulati2020conformer}. 
Whisper is a Transformer-based ASR model developed by OpenAI, 
trained on approximately 680,000 hours of audio data. 
Owing to its strong robustness and competitive recognition performance across diverse speech conditions, Whisper has become a widely adopted backbone for ASR research. 
In this work, we adopt Whisper-large-v3 as the target model for EAGLE-ASR. 
This model consists of a 32-layer Transformer encoder and a 32-layer Transformer decoder, with approximately 1550M parameters in total.
Conformer AED is a hybrid encoder-decoder model based on the CTC/attention architecture. In this work, we adopt the Conformer AED implementation provided by the FunASR toolkit\cite{gao2023funasr} as the target model for EAGLE-ASR. This model is trained on approximately 50,000 hours of English audio data and comprises a 32-layer Conformer encoder and a 16-layer Transformer decoder, with approximately 220M parameters in total. Since our study focuses on autoregressive decoding, the CTC branch is not used in inference.

\subsection{Training Details}
\label{training}
For the Whisper model, Speech-Draft is trained for 200K and 400K steps on the LibriSpeech and VoxPopuli datasets, respectively.
For the Conformer AED model, Speech-Draft is trained for 150K steps on LibriSpeech and 300K steps on VoxPopuli. 
Across all experiments, we employed the AdamW optimizer\cite{loshchilov2017decoupled} with beta values $(\beta_1, \beta_2) = (0.9, 0.95)$, together with a gradient clipping of $0.5$. 
A cosine learning rate schedule is adopted with a peak learning rate of $5 \times 10^{-5}$ and a warmup ratio of $0.05$\cite{loshchilov2016sgdr}. The batch size is set to $2$ with $4$ gradient accumulation steps, yielding an effective batch size of $8$. For the multi-step generation training strategy, we adopt a weight decay factor $\beta = 0.8$ and prediction depth $K = 7$.
All evaluations were conducted on a single NVIDIA RTX 4090 GPU with 24\,GB of memory.

\subsection{Evaluation Metrics}
\label{metrics}
Since EAGLE-ASR does not modify the parameters of the target model and adopts strict speculative decoding acceptance conditions\cite{xia2024unlocking}, \cite{xia2023speculative}, its decoded outputs are theoretically identical to those produced by the original target model. 
To empirically verify this lossless property in practice, we report $\Delta$WER, which measures the change in WER of EAGLE-ASR relative to the original target model. 
A $\Delta$WER greater than $0$ indicates that the decoding method incurs a degradation in recognition accuracy. Similarly, for ST tasks, we report the change in BLEU score\cite{papineni2002bleu}, denoted as $\Delta$BLEU, to measure the change in translation quality relative to the original model.

In addition to $\Delta$WER, we further adopt the following metrics to evaluate the acceleration performance:

(1) Speedup ratio: The relative inference speedup compared to the baseline (vanilla autoregressive decoding).

(2) Mean accepted length ($\tau$): The average number of draft tokens accepted during each verification step, which reflects the effectiveness of the draft generation process and is directly related to the achievable acceleration.


\section{Results and Analysis}
\label{sec:result}
\subsection{Main Results}
\label{main-result}
In this section, we evaluate the performance of EAGLE-ASR and other speculative decoding methods on the LibriSpeech benchmark using both Whisper and Conformer AED models. 
To ensure a fair comparison, all methods are re-implemented and evaluated under the same hardware configuration and experimental environment. 
Unless otherwise specified, the decoding hyperparameters for evaluation are set to a draft length of $K=5$ and $\text{top-}k=1$. 
For the standard Speculative Decoding (SpD) baselines with Whisper-large-v3, we employ Distil-large-v3 and Whisper-tiny as draft models. For the Conformer AED model, we use Distil-small and Whisper-tiny to ensure the draft model remains substantially smaller than the target.\footnote{Due to the discrepancy between the tokenizers of Conformer AED and Whisper, direct token verification is infeasible. We address this vocabulary mismatch by decoding the Whisper draft tokens into text and then re-encoding them using the Conformer tokenizer before verification.}
Regarding the Medusa baseline, we evaluate two implementations from Whisper-Medusa on the Whisper model: Medusa-Linear and Medusa-Block. It is crucial to note that neither of these approaches achieves lossless acceleration, as they modify the original parameters of the target model and do not adhere to strict speculative decoding acceptance criteria. 
In addition, we do not report Medusa's acceleration results on the Conformer AED model as no public implementation is currently available.

\begin{table*}[htbp]
\centering
\caption{Comparison of acceleration performance across different speculative decoding approaches on the LibriSpeech dataset.}
\label{tab:main_results}
\renewcommand{\arraystretch}{1.2} 
\resizebox{\linewidth}{!}{
\begin{tabular}{cccccccccc} 
\toprule
\multirow{2}{*}{\textbf{Target model}} & \multirow{2}{*}{\textbf{Method}} & \multirow{2}{*}{\textbf{Draft Model}} & \multirow{2}{*}{\textbf{Draft Params}} & \multicolumn{3}{c}{\textbf{Test Clean}} & \multicolumn{3}{c}{\textbf{Test Other}} \\
\cmidrule(lr){5-7} \cmidrule(lr){8-10} 
& & & & \textbf{Speedup} & \boldmath{$\tau$} & \textbf{$\Delta$WER} & \textbf{Speedup} & \boldmath{$\tau$} & \textbf{$\Delta$WER} \\
\midrule

    \multirow{5}{*}{\begin{tabular}{@{}c@{}} \textbf{Whisper} \\ (1550 M) \end{tabular}} 
    & SpD\cite{gandhi2023distil}            & Distil-large-v3 & 756 M & 1.72$\times$ & \textbf{4.14} & 0.00 & 1.52$\times$ & \textbf{3.76} & 0.00 \\
    & SpD\cite{gandhi2023distil}            & Whisper-tiny    & 39 M  & 1.76$\times$ & 3.04 & 0.00 & 1.58$\times$ & 2.56 & 0.00 \\
    & Medusa\cite{segal2025whisper} & Medusa-Linear   & 68 M & 1.67$\times$     & 3.51   & +1.88 & 1.59$\times$    & 3.32    & +3.61 \\
    & Medusa\cite{segal2025whisper} & Medusa-Block    & 92 M & 1.65$\times$     & 3.16    & +1.17 & 1.54$\times$    & 3.01    & +1.50 \\
    & EAGLE-ASR      & Speech-Draft    & 117 M & \textbf{1.91$\times$} & 3.79 & 0.00 & \textbf{1.79$\times$} & 3.55 & 0.00 \\
    \midrule
    
    \multirow{3}{*}{\begin{tabular}{@{}c@{}} \textbf{Conformer} \\ (220 M) \end{tabular}} 
    & SpD\cite{gandhi2023distil}       & Distil-small   & 166 M & 1.18$\times$ & 2.85 & 0.00 & 0.90$\times$ & 2.35 & 0.00 \\
    & SpD\cite{gandhi2023distil}      & Whisper-tiny   & 39 M  & 1.13$\times$ & 2.56 & 0.00 & 0.88$\times$ & 1.97 & 0.00 \\
    & EAGLE-ASR & Speech-Draft   & 10 M & \textbf{2.43$\times$} & \textbf{4.98} & 0.00 & \textbf{2.30$\times$} & \textbf{4.50} & 0.00 \\
    \bottomrule
\end{tabular}
} 
\end{table*}

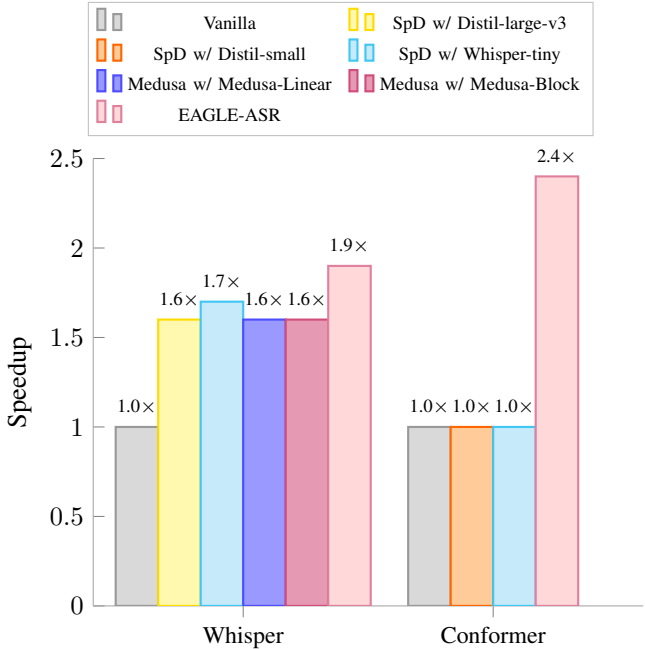
\begin{figure}[htbp]
    \centering
    \begin{tikzpicture}
    \begin{axis}[
        width=\linewidth,         
        height=7.5cm,             
        ybar=0pt,                 
        bar width=16pt,           
        ylabel={Speedup},
        ymin=0, 
        ymax=2.5,                 
        ytick={0,0.5,1,1.5,2,2.5},
        symbolic x coords={Whisper, Conformer},
        xtick={Whisper, Conformer},
        x tick label style={font=\small, align=center},
        enlarge x limits=0.6,    
        legend style={
            at={(0.45, 1.05)},     
            anchor=south,
            legend columns=2,     
            font=\scriptsize,     
            draw=gray!50,         
            fill=white,
            cells={align=left},
            /tikz/every even column/.append style={column sep=0.15cm}, 
        },
        nodes near coords,
        every node near coord/.append style={
            font=\scriptsize, 
            anchor=south,
            yshift=1pt            
        },
        point meta=explicit symbolic,
        axis lines*=left,
        axis line style={gray!70},
    ]
    
    \addlegendimage{ybar, fill=gray!30, draw=gray!80, thick}
    \addlegendentry{Vanilla}
    
    \addlegendimage{ybar, fill=yellow!40, draw=yellow!80!orange, thick}
    \addlegendentry{SpD w/ Distil-large-v3}
    
    \addlegendimage{ybar, fill=orange!40, draw=orange!80!red, thick}
    \addlegendentry{SpD w/ Distil-small}
    
    \addlegendimage{ybar, fill=cyan!20, draw=cyan!60, thick}
    \addlegendentry{SpD w/ Whisper-tiny}
    
    \addlegendimage{ybar, fill=blue!40, draw=blue!70, thick}
    \addlegendentry{Medusa w/ Medusa-Linear}
    
    \addlegendimage{ybar, fill=purple!40, draw=purple!70, thick}
    \addlegendentry{Medusa w/ Medusa-Block}
    
    \addlegendimage{ybar, fill=pink!60, draw=purple!50, thick}
    \addlegendentry{EAGLE-ASR}

    \addplot[fill=gray!30, draw=gray!80, thick, bar shift=-40pt, forget plot] coordinates {
        (Whisper, 1.0) [1.0$\times$]
    };
    \addplot[fill=yellow!40, draw=yellow!80!orange, thick, bar shift=-24pt, forget plot] coordinates {
        (Whisper, 1.6) [1.6$\times$]
    };
    \addplot[fill=cyan!20, draw=cyan!60, thick, bar shift=-8pt, forget plot] coordinates {
        (Whisper, 1.7) [1.7$\times$]
    };
    \addplot[fill=blue!40, draw=blue!70, thick, bar shift=8pt, forget plot] coordinates {
        (Whisper, 1.6) [1.6$\times$]
    };
    \addplot[fill=purple!40, draw=purple!70, thick, bar shift=24pt, forget plot] coordinates {
        (Whisper, 1.6) [1.6$\times$] 
    };
    \addplot[fill=pink!60, draw=purple!50, thick, bar shift=40pt, forget plot] coordinates {
        (Whisper, 1.9) [1.9$\times$]
    };

    \addplot[fill=gray!30, draw=gray!80, thick, bar shift=-24pt, forget plot] coordinates {
        (Conformer, 1.0) [1.0$\times$]
    };
    \addplot[fill=orange!40, draw=orange!80!red, thick, bar shift=-8pt, forget plot] coordinates {
        (Conformer, 1.0) [1.0$\times$]
    };
    \addplot[fill=cyan!20, draw=cyan!60, thick, bar shift=8pt, forget plot] coordinates {
        (Conformer, 1.0) [1.0$\times$]
    };
    \addplot[fill=pink!60, draw=purple!50, thick, bar shift=24pt, forget plot] coordinates {
        (Conformer, 2.4) [2.4$\times$]
    };

    \end{axis}
    \end{tikzpicture}
    \caption{Speedup comparison across different decoding methods. EAGLE-ASR consistently outperforms all baseline methods for both target models while guaranteeing lossless decoding.}
    \label{fig:speedup_bars}
\end{figure}

We present the results in \autoref{tab:main_results} and \autoref{fig:speedup_bars}.
For the Whisper-large-v3 target model, EAGLE-ASR achieves the highest speedup, reaching 1.91$\times$ on test-clean and 1.79$\times$ on test-other, outperforming all baselines while maintaining strictly lossless decoding ($\Delta\text{WER}=0.00$). 
Notably, SpD with Distil-large-v3 attains the highest acceptance length, yet its speedup ratio is lower than that of our EAGLE-ASR. 
This is due to the fact that the Distil-large-v3 draft model contains 756M parameters, comprising nearly half the size of the target model, which incurs substantial computational overhead during the drafting stage and consequently offsets the benefits gained from its high acceptance rate. In contrast, while other speculative decoding methods employ more lightweight draft models compared to EAGLE-ASR, the mean accepted lengths $\tau$ values are significantly lower than those of EAGLE-ASR, thus resulting in lower speedup ratios.
These results suggest that EAGLE-ASR achieves a more effective balance between draft quality and computational efficiency.

The advantage of EAGLE-ASR is more pronounced on the Conformer AED model. EAGLE-ASR achieves a remarkable 2.43$\times$ speedup on test-clean and 2.30$\times$ on test-other, substantially outperforming SpD with Distil-small and SpD with Whisper-tiny. In fact, both SpD baselines fail to accelerate, even resulting in a slowdown on the more challenging test-other set.
This poor performance can be attributed to a combination of factors.
First, the Conformer AED target model is relatively small (220M parameters), which makes the computational overhead of executing a separate draft model comparatively high. Second, a significant distributional mismatch exists between the Conformer AED target and the Whisper-based draft models. This includes discrepancies in transcription style, leading to low acceptance rates for the draft sequences (evidenced by low $\tau$ values for both SpD methods). Finally, due to the tokenizer mismatch between Conformer AED and Whisper models, we are compelled to perform token mapping through ground-truth text during the verification phase, which further exacerbates the time overhead. 
These results highlight the practical challenges of applying conventional speculative decoding across heterogeneous ASR model families.

Across both target models and evaluation sets, EAGLE-ASR consistently achieves the highest speedup while maintaining lossless transcription quality. These results highlight two key strengths of our proposed framework.
First, our proposed Speech-Draft module effectively leverages the hidden features of the target model to generate high-quality draft tokens while maintaining a lightweight architecture.
This design achieves an effective trade-off between generation quality and computational cost, ultimately leading to superior acceleration performance. 
Second, EAGLE-ASR demonstrates strong generalizability across target models with different architectures and parameter scales without requiring architecture-specific modifications or same-family distilled draft models. 
This makes EAGLE-ASR a practical and flexible acceleration solution for a broad range of autoregressive attention-based ASR systems, particularly for models that lack the same-family smaller variants or distilled lightweight variants.

\subsection{Performance on Multilingual Speech Recognition}
\label{multilingual-speech-recognition}
To further evaluate the generalizability of EAGLE-ASR across diverse languages, we report the acceleration performance of EAGLE-ASR on the VoxPopuli multilingual ASR benchmark using Whisper as the target model. 
Specifically, the evaluation covers five languages: English, German, Spanish, French, and Italian. 
The decoding hyperparameters for the multilingual ASR task remain identical to 
those used in the monolingual ASR experiments as described above.

\begin{table}[htbp]
    \centering
    \caption{Acceleration performance of EAGLE-ASR across diverse languages for ASR.}
    \label{tab:multilingual_speedup}
    \begin{tabular}{lccc}
        \toprule
        \textbf{Subset} & \textbf{Speedup} & \textbf{$\tau$} & \textbf{$\Delta $WER} \\
        \midrule
        VoxPopuli English & 1.96$\times$ & 4.06 & 0.00 \\
        VoxPopuli German  & 2.14$\times$ & 4.50 & 0.00 \\
        VoxPopuli Spanish & 2.10$\times$ & 4.46 & 0.00 \\
        VoxPopuli French  & 2.15$\times$ & 4.64 & 0.00 \\
        VoxPopuli Italian & 2.10$\times$ & 4.46 & 0.00 \\
        \midrule
        Average & 2.09$\times$ & 4.42 & 0.00 \\
        \bottomrule
    \end{tabular}
\end{table}

As shown in \autoref{tab:multilingual_speedup}, EAGLE-ASR delivers consistent and substantial acceleration across all five languages, achieving a 2.09$\times$ average acceleration while remaining strictly lossless ($\Delta$WER$=0.00$ for every subset). This performance is underpinned by the high quality of the generated draft tokens, as evidenced by the consistently high average acceptance lengths ($\tau$), which average 4.41 across the languages.
Notably, EAGLE-ASR exhibits highly consistent performance across different languages, with speedup ratios ranging from 1.96$\times$ to 2.15$\times$. 
The relatively small performance variation suggests that the Speech-Draft module effectively captures language-agnostic patterns from the target model's hidden representations, enabling it to generalize well across diverse languages.
Overall, these multilingual evaluations confirm that the acceleration benefits of EAGLE-ASR extend well beyond English, confirming its strong generalizability across diverse languages while strictly preserving lossless decoding accuracy.

\subsection{Performance on Speech Translation}
\label{translation}
To further evaluate the generalization ability and robustness of EAGLE-ASR beyond ASR, we extend our investigation to the speech translation (ST) task.
Specifically, by leveraging Whisper's inherent capability to directly translate non-English speech into English text, we conduct additional experiments on the VoxPopuli dataset. 
Unlike its earlier use in multilingual ASR, VoxPopuli is repurposed in this section to benchmark the acceleration performance of EAGLE-ASR across four diverse X-to-English translation directions. 
For the ST task, we report the acceleration results with a draft length of $K=5$ and a sampling size of $\text{top-}k=2$, which helps accommodate the increased output uncertainty compared with standard transcription task.

\begin{table}[htbp]
    \centering
    \caption{Acceleration performance of EAGLE-ASR across various speech translation tasks.}
    \label{tab:st_speedup}
    \begin{tabular}{lccc}
        \toprule
        \textbf{ST task} & \textbf{Speedup} & \textbf{$\tau$} & \textbf{$\Delta $BLEU}\\
        \midrule
        German-to-English & 1.49$\times$ & 3.44 & 0.00 \\
        Spanish-to-English & 1.73$\times$ & 3.75 & 0.00 \\
        French-to-English & 1.66$\times$ & 3.60 & 0.00 \\
        Italian-to-English  & 1.59$\times$ & 3.49 & 0.00 \\
        \midrule
        Average & 1.62$\times$ & 3.57 & 0.00 \\
        \bottomrule
    \end{tabular}
\end{table}

As shown in \autoref{tab:st_speedup}, EAGLE-ASR consistently improves inference efficiency across all four translation tasks, achieving an average speedup of 1.62$\times$ and a mean acceptance length $\tau$ of 3.57. 
Importantly, this acceleration remains strictly lossless, with a $\Delta$BLEU score of 0.00 for every task, confirming that EAGLE-ASR fully preserves the translation quality of the original Whisper model.
A notable observation is that, despite using a larger $\text{top-}k$ value, EAGLE-ASR achieves a lower mean accepted length on speech translation ($\tau=3.57$) than on multilingual ASR task ($\tau=4.42$). 
This behavior is expected due to the inherently higher complexity of the ST task.
Unlike ASR, ST involves a more intricate cross-lingual mapping from source speech to target text, where the increased output uncertainty poses greater challenges for draft model predictions to be accepted. 
As a result, fewer draft tokens are accepted during verification. Together with the additional computational overhead introduced by the larger top-k setting, this explains the relatively lower speedup observed for the ST task (1.62$\times$) compared with multilingual ASR task (2.09$\times$).
Nevertheless, despite the increased task difficulty, EAGLE-ASR demonstrates remarkable robustness and effectiveness by delivering substantial, lossless acceleration for speech translation. These results indicate that the benefits of EAGLE-ASR extend beyond monolingual and multilingual ASR and transfer effectively to challenging speech translation scenarios. 
This consistent acceleration across diverse cross-lingual generation tasks further underscores the strong generalization ability and robustness of our proposed framework.

\subsection{Impact of Decoding Hyperparameters}
\label{hyperparameters}
In this section, we investigate the impact of decoding hyperparameters on the acceleration performance of EAGLE-ASR. 
Specifically, we focus on two key parameters: the draft length $K$, which controls the number of tokens generated by the draft model at each speculative step, and the $\text{top-}k$ sampling size, which determines the number of candidate tokens considered during the sampling process.

\begin{figure}[t]
    \centering
    \begin{tikzpicture}
        \begin{axis}[
            width=0.26\textwidth,
            height=0.26\textwidth,
            xlabel={Draft Length $K$},
            ylabel={Speedup ($\times$)},
            xtick={2,3,4,5,6,7,8},
            xticklabels={2,3,4,5,6,7,8},
            ymin=1.5, ymax=2.0,
            ytick={1.5, 1.6,1.7,1.8,1.9,2.0},
            grid=major,
            grid style={dashed, gray!30},
            legend style={at={(0.98,0.98)}, anchor=north east, font=\small},
            tick label style={font=\small},
            label style={font=\small},
        ]
        \addplot[color=orange, mark=*, mark size=2pt, thick,
                 mark options={solid, fill=orange}]
            coordinates {
                (2,1.64)(3,1.76)(4,1.83)(5,1.85)
                (6,1.84)(7,1.83)(8,1.63)
            };
        \end{axis}
    \end{tikzpicture}
    \hfill
    \begin{tikzpicture}
        \begin{axis}[
            width=0.26\textwidth,
            height=0.26\textwidth,
            xlabel={Draft Length $K$},
            ylabel={$\tau$},
            xtick={2,3,4,5,6,7,8},
            xticklabels={2,3,4,5,6,7,8},
            ymin=1, ymax=5,
            ytick={1,2,3,4,5},
            grid=major,
            grid style={dashed, gray!30},
            legend style={at={(0.98,0.98)}, anchor=north east, font=\small},
            tick label style={font=\small},
            label style={font=\small},
        ]
        \addplot[color=orange, mark=*, mark size=2pt, thick,
                 mark options={solid, fill=orange}]
            coordinates {
                (2,1.82)(3,2.66)(4,3.29)(5,3.67)
                (6,4.04)(7,4.35)(8,4.57)
            };
        \end{axis}
    \end{tikzpicture}%
    \caption{Effect of the draft length $K$ on the acceleration performance of EAGLE-ASR. Left: Speedup ratio. Right: Mean accepted length $\tau$.}
    \label{fig:draft length}
\end{figure}
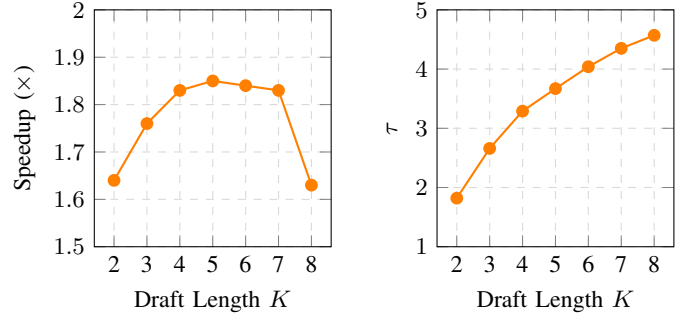

\autoref{fig:draft length} illustrates how the draft length $K$ affects both the speedup ratio and the mean accepted length $\tau$ when $\text{top-}k$ is fixed to 1.
As shown in the right panel, the mean accepted length $\tau$ increases monotonically as $K$ increases, rising from 1.82 at $K{=}2$ to 4.57 at $K{=}8$. 
This is expected, since a longer draft sequence provides more candidate tokens for verification, and thus more tokens can be accepted on average in a single forward pass.

However, the speedup ratio does not exhibit the same monotonic trend. 
As shown in the left panel, the speedup first increases with $K$, reaches a peak of approximately 1.85$\times$ at $K{=}5$, and then gradually declines as $K{=}5$ increases further.
This discrepancy can be explained by the trade-off between two competing factors: the number of accepted tokens and the computational cost of drafting. On the one hand, a larger $K$ leads to a higher $\tau$, meaning that fewer verification steps are needed to generate a complete sequence, thereby reducing the number of expensive forward passes through the target model. On the other hand, each draft step requires the draft model to run $K$ autoregressive iterations, and the subsequent verification pass must process the entire draft sequence, both of which incur additional computational overhead that scales roughly with $K$.
When $K$ is small, the gain from accepting more tokens per step outweighs the extra drafting cost, leading to improved acceleration. 
However, the marginal gain in accepted tokens gradually diminishes as $K$ grows.
For example, $\tau$ increases by 0.84 when $K$ goes from 2 to 3, but only by 0.22 when $K$ increases from 7 to 8. 
This sublinear growth of $\tau$ suggests that draft tokens predicted further are more likely to be rejected due to error accumulation. 
Beyond the optimal point ($K=5$), the additional drafting overhead outweighs the modest gain in accepted length, leading to reduced overall speedup.


\begin{figure}[t]
    \centering
    \begin{tikzpicture}
        \begin{axis}[
            width=0.26\textwidth,
            height=0.26\textwidth,
            xlabel={$\text{top-}k$},
            ylabel={Speedup ($\times$)},
            xtick={1,2,3,4},
            xticklabels={1,2,3,4},
            ymin=1.5, ymax=1.9,
            ytick={1.5, 1.6,1.7,1.8,1.9},
            grid=major,
            grid style={dashed, gray!30},
            legend style={at={(0.98,0.98)}, anchor=north east, font=\small},
            tick label style={font=\small},
            label style={font=\small},
        ]
        \addplot[color=orange, mark=*, mark size=2pt, thick,
                 mark options={solid, fill=orange}]
            coordinates {
                (1,1.85)(2,1.80)(3,1.70)(4,1.54)
            };
        \end{axis}
    \end{tikzpicture}
    \hfill
    \begin{tikzpicture}
        \begin{axis}[
            width=0.26\textwidth,
            height=0.26\textwidth,
            xlabel={$\text{top-}k$},
            ylabel={$\tau$},
            xtick={1,2,3,4},
            xticklabels={1,2,3,4},
            ymin=3, ymax=5,
            ytick={3,3.5,4,4.5,5},
            grid=major,
            grid style={dashed, gray!30},
            legend style={at={(0.98,0.98)}, anchor=north east, font=\small},
            tick label style={font=\small},
            label style={font=\small},
        ]
        \addplot[color=orange, mark=*, mark size=2pt, thick,
                 mark options={solid, fill=orange}]
            coordinates {
                (1,3.67)(2,4.13)(3,4.42)(4,4.56)
            };
        \end{axis}
    \end{tikzpicture}%
    \caption{Effect of $\text{top-}k$ on the acceleration performance of EAGLE-ASR. Left: Speedup ratio. Right: Mean accepted length $\tau$.}
    \label{fig:top-k}
\end{figure}

\autoref{fig:top-k} further presents the impact of the draft $\text{top-}k$ when the draft length is fixed at $K{=}5$.
As shown in the right panel, the mean accepted length $\tau$ increases monotonically as $\text{top-}k$ increases, rising from 3.67 at $\text{top-}k{=}1$ to 4.56 at $\text{top-}k{=}4$. 
This trend is expected, since enlarging the candidate set at each drafting step increases the likelihood that the target token is included among the proposed candidates, thereby improving the acceptance rate during verification.
However, the speedup ratio exhibits the opposite trend.
As shown in the left panel, the speedup decreases monotonically from 1.85$\times$ at $\text{top-}k{=}1$ to 1.54$\times$ at $\text{top-}k{=}4$. 
This indicates that, for ASR tasks, EAGLE-ASR achieves greater acceleration when adopting linear verification ($\text{top-}k = 1$) compared to tree-based verification ($\text{top-}k > 1$). 
Notably, this finding contrasts with prior empirical observations in natural language processing, where speculative decoding typically benefits more from tree-based verification strategies.
We attribute this phenomenon to the simplicity and deterministic nature of ASR tasks. As $\text{top-}k$ increases, the gain in the accepted token length $\tau$ is marginal. For example, when $\text{top-}k$ increases from 1 to 2, $\tau$ improves by only 0.46. Concurrently, a larger $\text{top-}k$ leads the draft model to generate more candidate tokens at each step, thereby increasing the number of tokens that must be verified by the target model and consequently raising the computational cost per verification step. The modest improvement in $\tau$ brought by increasing $\text{top-}k$ is insufficient to offset the steadily growing verification overhead, ultimately resulting in a consistent decline in the overall speedup ratio.

To validate this hypothesis, we further investigate the impact of $\text{top-}k$ on the acceleration performance of EAGLE-ASR in the ST task. Compared with ASR, ST is inherently more complex and exhibits greater output diversity, as a single source utterance can have multiple valid translations. We present the results for the German-to-English task as a representative example.

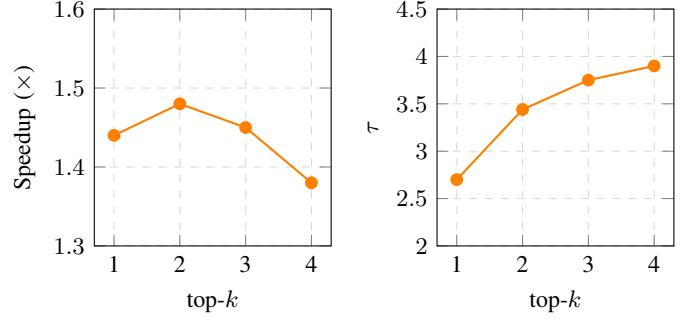
\begin{figure}[t]
    \centering
    \begin{tikzpicture}
        \begin{axis}[
            width=0.26\textwidth,
            height=0.26\textwidth,
            xlabel={$\text{top-}k$},
            ylabel={Speedup ($\times$)},
            xtick={1,2,3,4},
            xticklabels={1,2,3,4},
            ymin=1.3, ymax=1.6,
            ytick={1.3, 1.4,1.5,1.6},
            grid=major,
            grid style={dashed, gray!30},
            legend style={at={(0.98,0.98)}, anchor=north east, font=\small},
            tick label style={font=\small},
            label style={font=\small},
        ]
        \addplot[color=orange, mark=*, mark size=2pt, thick,
                 mark options={solid, fill=orange}]
            coordinates {
                (1,1.44)(2,1.48)(3,1.45)(4,1.38)
            };
        \end{axis}
    \end{tikzpicture}
    \hfill
    \begin{tikzpicture}
        \begin{axis}[
            width=0.26\textwidth,
            height=0.26\textwidth,
            xlabel={$\text{top-}k$},
            ylabel={$\tau$},
            xtick={1,2,3,4},
            xticklabels={1,2,3,4},
            ymin=2, ymax=4.5,
            ytick={2,2.5,3,3.5,4,4.5},
            grid=major,
            grid style={dashed, gray!30},
            legend style={at={(0.98,0.98)}, anchor=north east, font=\small},
            tick label style={font=\small},
            label style={font=\small},
        ]
        \addplot[color=orange, mark=*, mark size=2pt, thick,
                 mark options={solid, fill=orange}]
            coordinates {
                (1,2.70)(2,3.44)(3,3.75)(4,3.90)
            };
        \end{axis}
    \end{tikzpicture}%
    \caption{Effect of $\text{top-}k$ on the acceleration performance of EAGLE-ASR on the ST (German-to-English) task. Left: Speedup ratio. Right: Mean accepted length $\tau$.}
    \label{fig:topk-st}
\end{figure}

\autoref{fig:topk-st} shows the effect of $\text{top-}k$ on EAGLE-ASR for the ST task. In contrast to the ASR results, the speedup on ST does not decrease monotonically as $\text{top-}k$ increases. Instead, the speedup first improves from 1.44$\times$ at $\text{top-}k{=}1$ to 1.48$\times$ at $\text{top-}k{=}2$, and then gradually declines to 1.38$\times$ at $\text{top-}k{=}4$. Meanwhile, the mean accepted length $\tau$ increases steadily from 2.70 to 3.90.
This result supports our hypothesis that the effectiveness of tree-based verification is task dependent. For the ST task, increasing the $\text{top-}k$ value enables the draft model to explore a broader range of translation hypotheses, thereby increasing the likelihood that the target model accepts at least one branch within the expanded tree. Consequently, the growth in $\tau$ induced by larger $\text{top-}k$ is substantially more pronounced in ST than in ASR. In particular, when $\text{top-}k$ increases from 1 to 2, $\tau$ rises by 0.74, which is substantially higher than the corresponding gain of 0.46 observed in the ASR task. This improvement is sufficient to offset the additional verification overhead, thereby yielding a speedup.
However, the benefit quickly saturates when $\text{top-}k$ becomes larger. When $\text{top-}k$ exceeds 2, the verification overhead begins to dominate: the marginal increase in $\tau$ diminishes, while the number of candidates requiring verification continues to grow, leading to degraded speedup. We conjecture that this is because ST tasks, despite being more complex than ASR, remain relatively simple and deterministic compared to dialogue or comprehension tasks, which typically employ $\text{top-}k = 4$ or $8$ in speculative decoding settings\cite{li2024eagle2}, \cite{cai2024medusa}.


\subsection{Ablation Studies}
\label{ablation}
In this section, we conduct a comprehensive ablation study to systematically evaluate the contribution of each component in EAGLE-ASR. Our analysis focuses on three primary aspects: the architecture of the Speech-Draft module, the training data, and the employed training strategies.

\begin{table}[htbp]
    \centering
    \caption{Ablation study results of EAGLE-ASR on the Whisper model.}
    \label{tab:abalation}
    \begin{tabular}{lccc}
        \toprule
        \textbf{Method} & \textbf{Speedup} & \textbf{$\tau$} & \textbf{$\Delta $WER}\\
        \midrule
        w/o encoder feature & 0.89$\times$ & 0.08 & 0.00 \\
        w/o fused decoder features & 1.35$\times$ & 1.84 & 0.00 \\
        \midrule
        w/o self-distilled data & 1.43$\times$ & 2.02 & 0.00 \\
        \midrule
        w/o multi-step generation & 1.17$\times$ & 1.00 & 0.00 \\
        w/o ${L}_{\text{mtp}}$ & 1.66$\times$ & 3.28 & 0.00 \\
        \midrule
        EAGLE-ASR  & 1.85$\times$ & 3.67 & 0.00 \\
        \bottomrule
    \end{tabular}
\end{table}

We first investigate the impact of different architectural components within the Speech-Draft module. As shown in the first two rows of \autoref{tab:abalation}, both encoder features and fused decoder features play critical roles in achieving effective speculative decoding.
When the encoder features are removed by replacing the cross-attention mechanism with self-attention, the model suffers from catastrophic performance degradation. Instead of achieving acceleration, the system exhibits a slowdown (0.89$\times$) and a drastically reduced mean accepted length ($\tau = 0.08$). This indicates that the acoustic information from the encoder is indispensable for the Speech-Draft module to generate plausible draft tokens that align with the target ASR model's predictions. Without direct access to encoder representations, the draft model lacks sufficient acoustic grounding, leading to poor-quality predictions that are rarely accepted by the target model.

Next, when replacing the fused decoder features with only the second-top-layer decoder features, the model still achieves acceleration (1.35$\times$ speedup), but the improvement is substantially lower than that of the full EAGLE-ASR framework. The results demonstrate that the fusion of multi-layer decoder features provides richer contextual representations. By integrating information from different levels of the decoding process, the fused features enable the Speech-Draft module to more accurately approximate the behavior of the target model and to generate higher-quality draft tokens, thereby improving the overall acceleration performance.

Then we analyze the effect of the training data. The third row of \autoref{tab:abalation} presents the results when the Speech-Draft module is trained on the original ground-truth transcripts rather than on self-distilled data generated by the target ASR model. This change leads to a noticeable degradation in acceleration performance, with the speedup dropping to 1.43$\times$ and the mean accepted length $\tau$ falling to 2.02. 
This result highlights a fundamental principle of speculative decoding: the draft model's primary goal is to predict the target model's output, not necessarily the ground truth. When trained on self-distilled data, the Speech-Draft module directly learns to approximate the token distribution produced by the target ASR model, including transcription style characteristics such as punctuation and capitalization. Such distributional alignment increases the likelihood that the drafted tokens will be accepted during the verification stage. 
In contrast, training on ground-truth transcripts forces the draft model to learn an ``ideal'' distribution, which may diverge from the target model's. This distributional mismatch causes more rejections, as predictions that are correct with respect to the ground truth may nonetheless differ from the tokens that the target model would generate. Moreover, in the case of the Whisper model, the original transcriptions lack special token prefixes, which further constrains the acceptance rate of draft tokens generated by the Speech-Draft at the initial decoding step.

Finally, we examine the impact of our proposed training strategies. The fourth and fifth rows of \autoref{tab:abalation} show the results of ablating the multi-step generation training and the weighted cross-entropy loss with exponential decay $L_{\text{mtp}}$.
When the number of iterative generation steps during training is set to one, a dramatic performance collapse is observed, with the speedup plummeting to $1.17\times$ and the mean accepted length $\tau$ reducing to 1.00. The result of $\tau = 1.00$ indicates that, on average, only the first token of each drafted sequence is accepted. This outcome aligns with our expectations: during training, Speech-Draft is optimized solely to generate predictions conditioned on the hidden representations of the target model, rather than on its own previously predicted hidden states. Consequently, this module is only capable of predicting the immediate next token and lacks the ability to autoregressively generate subsequent tokens. This observation highlights that a multi-step generation training strategy is essential for enabling Speech-Draft to produce multiple plausible tokens simultaneously, thereby enhancing its effectiveness in speculative decoding.
When the exponential decay weighting in $L_{\text{mtp}}$ is removed (setting $\beta = 1$), the model achieves a speedup of 1.66$\times$ with $\tau = 3.28$. Although this still provides a noticeable acceleration, it remains inferior to the full EAGLE-ASR configuration. This degradation occurs because predicting tokens further into the future is inherently more challenging due to increased uncertainty. The exponential decay mechanism in ${L}_{\text{mtp}}$ strategically prioritizes the accuracy of the initial tokens in the drafted sequence, which are more critical for the acceptance of the entire sequence. By assigning equal weights across all positions, the model is forced to allocate its capacity to optimizing for harder, more distant predictions, potentially at the cost of accuracy on the more crucial initial tokens. Since an incorrect early token leads to the rejection of the entire draft, this targeted optimization leads to improved draft quality and higher acceptance rates during verification. Since an error early token leads to the rejection of the entire draft sequence, uniform weighting can ultimately lead to a lower overall acceptance rate.

\section{Conclusion}
\label{sec:conclude}
In this paper, we propose EAGLE-ASR, a novel and general framework for lossless inference acceleration of autoregressive ASR models through EAGLE-style speculative decoding. 
By introducing a lightweight and model-agnostic Speech-Draft module that jointly leverages hierarchical decoder representations and encoder acoustic features, the proposed framework effectively balances draft quality and computational overhead. 
Through self-distilled training data and a multi-step generation strategy, Speech-Draft is optimized to closely approximate the output distribution of the target model while maintaining stable and reliable multi-token prediction capability. 
Extensive experiments conducted on Whisper and Conformer-based AED models demonstrate that EAGLE-ASR achieves substantial inference speedups. 
It consistently outperforms existing speculative decoding baselines, while strictly preserving output quality without any degradation.
Furthermore, additional experiments on multilingual ASR and speech translation tasks confirm the robustness and generalizability of EAGLE-ASR across diverse languages and cross-lingual generation settings.
Overall, EAGLE-ASR provides a practical and universally applicable solution for accelerating modern ASR systems without sacrificing accuracy.

\bibliographystyle{IEEEtran}
\bibliography{refs}

\vfill

\end{document}